\documentclass[aps,prl,final,10pt,twocolumn,showpacs,superscriptaddress,footinbib]{revtex4-1}
\usepackage{amsmath,amssymb}
\usepackage{graphicx,color}
\usepackage[squaren]{SIunits}
\usepackage[english]{babel}

\newcommand{\dif}{\mathrm{d}}

\begin{document}
\title{Reply to Comment on ``Circular Motion of Asymmetric Self-Propelling Particles''}

\author{Felix K{\"u}mmel}
\affiliation{2.\ Physikalisches Institut, Universit\"at Stuttgart, D-70569 Stuttgart, Germany} 

\author{Borge ten Hagen}
\affiliation{Institut f{\"u}r Theoretische Physik II: Weiche Materie,
Heinrich-Heine-Universit{\"a}t D{\"u}sseldorf, D-40225 D{\"u}sseldorf, Germany}

\author{Raphael Wittkowski}
\affiliation{SUPA, School of Physics and Astronomy, University of Edinburgh, Edinburgh, EH9 3JZ, United Kingdom}

\author{Daisuke Takagi}
\affiliation{Department of Mathematics, University of Hawaii at Manoa, Honolulu, Hawaii
96822, USA}

\author{Ivo Buttinoni}
\affiliation{2.\ Physikalisches Institut, Universit\"at Stuttgart, D-70569 Stuttgart, Germany} 

\author{Ralf~Eichhorn} 
\affiliation{Nordita, Royal Institute of Technology, and Stockholm  
University, SE-10691 Stockholm, Sweden} 

\author{Giovanni Volpe}
\affiliation{2.\ Physikalisches Institut, Universit\"at Stuttgart, D-70569 Stuttgart, Germany} 
\affiliation{Present address: Department of Physics, Bilkent University, Cankaya,
Ankara 06800, Turkey}

\author{Hartmut L{\"o}wen}
\affiliation{Institut f{\"u}r Theoretische Physik II: Weiche Materie,
Heinrich-Heine-Universit{\"a}t D{\"u}sseldorf, D-40225 D{\"u}sseldorf, Germany}

\author{Clemens Bechinger}
\affiliation{2.\ Physikalisches Institut, Universit\"at Stuttgart, D-70569 Stuttgart, Germany} 
\affiliation{Max-Planck-Institut f\"ur Intelligente Systeme, D-70569 Stuttgart, Germany}

\date{\today}


\maketitle


In a Comment \cite{CommentFelderhof} on our Letter on self-propelled asymmetric particles \cite{kummel2013circular}, 
Felderhof claims that our theory based on Langevin equations would be conceptually wrong. In this Reply we show that our theory is appropriate, consistent, and physically justified.

The motion of a self-propelled particle (SPP) is force-
and torque-free if external forces and torques are absent.
Nevertheless, as stated in our Letter \cite{kummel2013circular}, \textit{effective} forces and torques \cite{Friedrich2008,Jekely2008,Radtke2012,Crespi2013,Marine2013} can be used 
together with the grand resistance matrix (GRM) \cite{Kraft:13}
to describe the self-propulsion of force- and torque-free swimmers \footnote{Following common nomenclature, we do not distinguish between the terms ``self-propulsion'' and ``swimming'' with regard to the rigidity of the particle.}.
To prove this, we perform a hydrodynamic calculation based on slender-body theory for Stokes flow \cite{Batchelor,unpublished}. This approach has been applied successfully to model, e.g., flagellar locomotion \cite{Lighthill,Lauga_review:09} and avoids a general Fax\'en's theorem for asymmetric particles. A key assumption of slender-body theory is that the width $2\epsilon$ of the arms of the L-shaped particle is much smaller than the total arc length $L=a+b$, where $a$ and $b$ are the arm lengths.

The centerline position of the slender particle is 
$\mathbf{x}(s)=\mathbf{r}-\mathbf{r}_{\mathrm{S}}+s\mathbf{\hat{u}}_{\parallel}$ for
$-b\le s\le 0$ and $\mathbf{x}(s)=\mathbf{r}-\mathbf{r}_{\mathrm{S}}+s
\mathbf{\hat{u}}_{\perp}$ for $0< s\le a$. 
Here, $\mathbf{r}$ is the center-of-mass position of the particle in the laboratory frame of reference and $\mathbf{r}_{\mathrm{S}}=(a^2\mathbf{\hat{u}}_\perp-b^2\mathbf{\hat{u}}_\parallel)/(2L)$ is a vector in the particle's frame---defined by the unit vectors $\mathbf{\hat{u}}_\parallel$, $\mathbf{\hat{u}}_\perp$---such that $\mathbf{r}-\mathbf{r}_{\mathrm{S}}$ is the point where the two arms meet at right angles. The fluid velocity on the particle surface is approximated by $\mathbf{\dot{x}}+\mathbf{v}_\mathrm{sl}$ with a prescribed slip velocity $\mathbf{v}_\mathrm{sl}(s)$. According to the leading-order slender-body approximation \cite{Batchelor}, the fluid velocity is related to the local force per unit length $\mathbf{f}(s)$ on the particle surface by
$\mathbf{\dot{x}}+\mathbf{v}_\mathrm{sl}=c (\mathbf{I}+\mathbf{x'}\!\!\otimes\!\mathbf{x'})\mathbf{f}$ with $c=\log(L/\epsilon)/(4\pi\eta$), the solvent viscosity $\eta$, the identity matrix $\mathbf{I}$, $\mathbf{x'}=\partial\mathbf{x}/\partial s$, and the dyadic product $\otimes$. The force density $\mathbf{f}$ satisfies the integral constraints of vanishing net force,
$\int_{-b}^a\mathbf{f}\,\dif s=\mathbf{0}$,
and vanishing net torque relative to the center of mass,
$\mathbf{\hat{e}}_z\!\cdot\!\!\int_{-b}^a(-\mathbf{r}_{\mathrm{S}}+s\mathbf{x'})\!\times\!\mathbf{f}\,\dif s
=\!\!\int_{-b}^0 s\mathbf{\hat{u}}_\perp\!\!\cdot\!\mathbf{f}\,\dif s
-\!\int_{0}^a s\mathbf{\hat{u}}_\parallel\!\cdot\!\mathbf{f}\,\dif s=0$, with $\mathbf{\hat{e}}_z=(0,0,1)^\mathrm{T}$.

First, we consider a \textit{passive} particle driven by an external force $\mathbf{F}_\mathrm{ext}$, which is constant in the particle's frame, and torque $M_\mathrm{ext}$. For this case, we assume no-slip conditions for the fluid on the entire particle surface.
Then the integral constraints with net force $\mathbf{F}_\mathrm{ext}$ and torque $M_\mathrm{ext}$ give
\begin{eqnarray} 
\eta \boldsymbol{\mathcal{H}} \left(
\mathbf{\hat{u}}_\parallel\!\cdot\!\mathbf{\dot{r}}, 
\mathbf{\hat{u}}_\perp\!\!\cdot\!\mathbf{\dot{r}},
\dot{\phi}\right)^\mathrm{T}
=\left(\mathbf{\hat{u}}_\parallel\!\cdot\!\mathbf{F}_\mathrm{ext},
\mathbf{\hat{u}}_\perp\!\!\cdot\!\mathbf{F}_\mathrm{ext},
M_\mathrm{ext}\right)^\mathrm{T},
\label{eq:ext}%
\end{eqnarray}
where 
\begin{eqnarray}
\boldsymbol{\mathcal{H}} = \frac{1}{2c\eta}
\begin{pmatrix}
2a+b & 0 & -a^2b /(2L) \\
0 & a+2b & -ab^2/(2L) \\
-a^2b/(2L) & -ab^2/(2L) & A
\end{pmatrix}
\label{eq:H}%
\end{eqnarray}
with $A=((8L^2-3ab)(a^3+b^3)-6L(a^4+b^4))/(12L^2)$ is the GRM that depends on the particle shape \cite{HappelB1991,Kraft:13}. 

In the \textit{self-propelled} case, motivated by the slip flow generated near the Au coating in the experiments, we set $\mathbf{v}_\mathrm{sl}=-V_\mathrm{sl}\mathbf{\hat{u}}_\perp$ along the arm of length $b$ and no slip ($\mathbf{v}_\mathrm{sl}=\mathbf{0}$) along the other arm. This results in  
\begin{eqnarray}
\eta \boldsymbol{\mathcal{H}}
\left(
\mathbf{\hat{u}}_\parallel\!\cdot\!\mathbf{\dot{r}}, 
\mathbf{\hat{u}}_\perp\!\!\cdot\!\mathbf{\dot{r}}, 
\dot{\phi}\right)^\mathrm{T}
=\left(0, 
bV_\mathrm{sl}/c, 
-ab^2 V_\mathrm{sl}/(2cL)\right)^\mathrm{T}.\;\; 
\label{eq:sprop}%
\end{eqnarray}
We emphasize that the tensor $\boldsymbol{\mathcal{H}}$ in Eq.\ \eqref{eq:sprop} is identical to the GRM in Eq.\ \eqref{eq:ext}. Formally, both equations are \textit{exactly} the same if $\mathbf{\hat{u}}_\parallel\!\cdot\!\mathbf{F}_\mathrm{ext}=0$, $\mathbf{\hat{u}}_\perp\!\!\cdot\!\mathbf{F}_\mathrm{ext}=bV_\mathrm{sl}/c$, and $M_\mathrm{ext}=-ab^2 V_\mathrm{sl}/(2cL)$. This shows that the motion of a SPP with $\mathbf{v}_\mathrm{sl}=-V_\mathrm{sl}\mathbf{\hat{u}}_\perp$ along the arm of length $b$ is identical to the motion of a passive particle driven by a net external force $\mathbf{F}_\mathrm{ext}=F\mathbf{\hat{u}}_\perp$ and torque $M_\mathrm{ext}=lF$ with the effective self-propulsion force $F=bV_\mathrm{sl}/c$ and effective lever arm $l=-ab/(2L)$. 
By transforming Eq.\ \eqref{eq:sprop} from the particle's frame to the laboratory frame and introducing the generalized diffusion tensor 
$\boldsymbol{\mathcal{D}}=\boldsymbol{\mathcal{H}}^{-1}/(\beta\eta)$ \cite{unpublished}, where $\beta$ is the inverse effective thermal energy, one directly obtains the noise-free version of the equations of motion (EOMs) (1) in our Letter \cite{kummel2013circular}.

Clearly, for the same particle velocity, the flow and pressure fields generated by the SPP and the externally driven particle are different. However, the EOMs are the same. Therefore we can formally use external forces and torques that move with the SPP to model its self-propelled motion. In that sense, the concept of \textit{effective} forces and torques is justified, the application of the GRM is appropriate, and the EOMs in our Letter correctly describe
the dynamics of the SPP.

\end{document}